# Some Notes on the Rapanui Archaeoastronomy

Sergei Rjabchikov[1]

[1]The Sergei Rjabchikov Foundation - Research Centre for Studies of Ancient Civilisations and Cultures, Krasnodar, Russia, e-mail: srjabchikov@hotmail.com

**Abstract**

This paper is dedicated to the research of secrets of Easter Island (Rapa Nui), a remote plot of land in the Pacific; the work includes not only necessary ethnological data, but also some results on the archaeoastronomy. The analysis of several rock drawings lets us date them. The priests *Hina Mango* and *Rahu* (*Rahi*) were not only experts on the script, but also great astronomers. There is abundant evidence that the priests-astronomers used the astrolabe in their studies. The local astronomical terminology has been decoded. The observatory at the ceremonial platform Ahu Tongariki has been investigated carefully. The orientation of a female statue on the slope of the Rano Raraku volcano allows us to suggest that it was an image of the Moon (the moon goddess). A number of additional astronomical and calendar records in the rock art and in the writing have been deciphered.

**Astronomical Simulations as the Main Clue to the Mystery: The Dating of Archaic Petroglyphs**

I have investigated an archaic Rapanui rock drawing located near the ceremonial village of Orongo; besides, a fragment of the corresponding song has been translated.[2] The picture is well known;[3] see figure 1.

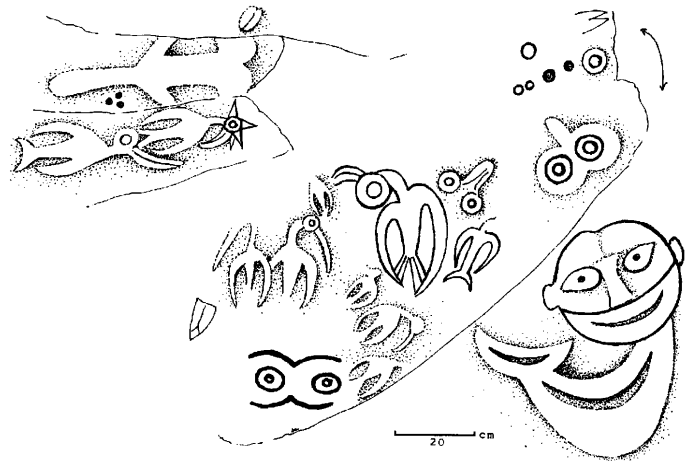

Figure 1.

One can try to date this scene. (Here and below I have used the computer program RedShift Multimedia Astronomy (Maris Multimedia, San Rafael, USA) to look at the heavens above Easter Island. The

---

[2] Rjabchikov, Sergei, 2013. The Astronomical and Ethnological Components of the Cult of Bird-Man on Easter Island. arXiv:1309.6056 [physics.hist-ph], 24 September, 2013, pp. 2-4.
[3] Lee, Georgia, 1992. *The Rock Art of Easter Island. Symbols of Power, Prayers to the Gods*. Los Angeles: The Institute of Archaeology Publications (UCLA), p. 102, figure 4.97.

following radio-carbon dates in Orongo and in the neighbourhood were obtained: Complex A: A.D. 1420 ± 70; Complex B (the village): A.D. 1416 ± 100, A.D. 1540 ± 100, A.D. 1576 ± 100.[4])

The key symbol is glyph **69** *moko* (lizard) represented on the left.[5] It is the designation of the night/day *Hiro*, the new moon in the local lunar calendar. The nearness of this glyph and glyph **1** *Tiki* (the sun deity, known as *Makemake*) can be interpreted as the message about a solar eclipse.

The calculations have been carried out for the 4th lunar month (beginning in September in the majority of instances) from A.D. 1300 till A.D 1700. As a result, the rock record reports concerning the partial solar eclipse of **August 29, A.D. 1467** (the 1st day, *Hiro* of the 4th month, *Hora-nui*). In this case the heliacal risings of β and α Centauri occurred on August 26 and September 4, A.D. 1467 respectively. Both stars are represented as a *rei-miro* pendant united with *Makemake*'s head (here it is the rising sun). The lizard sign and 3 dots (cupules), in the other words, 4 days as well as a row of 4 dots (cupules) near glyph **39** *raa* (the sun; day), i.e., 4 days, might be one and the same interval between the appearance of β Centauri and the eclipse.

This dating of the rock picture (after **August 29, A.D. 1467**) is vital. It is clear that the activity of the men at the Orongo area was permanent starting in ca. A.D. 1416.

It should be borne in mind that the iconography of two wooden statuettes of the bird-men, in St. Petersburg and New York, is directly concerned with the local petroglyphs at Orongo. Consider the next rock picture.[6] One of bird-men on the picture has glyph **4** *atua* (deity; lord) on his neck, and St. Petersburg specimen has the same glyph on the neck.[7] The bird-man in the rock has the sign of the face on his chest, and the New York specimen has glyph **60** *Mata* (Face; the symbol of the god *Tiki-Makemake*) on the chest.[8] The bird-man in the rock has glyph **1** *Tiki* on his belly, and the New York specimen has the same glyph on the belly. These data show once again that the bird-man was an incarnation of the warm and hot sun. This cult required astronomical observations of the sun, the moon and certain stars.

**The Observations of the Stars Aldebaran and Canopus: The Addition Information**

In my previous research three azimuths – 322.1°, 339.1° and 177.5° – of lines marked on the Mataveri stone calendar have been decoded: they bore on Aldebaran and Canopus.[9] Both stars were seen high in the sky, so that a simple astronomical device, astrolabe, was necessary for that purpose.

In the inscription on the Aruku-Kurenga tablet (B) an astrolabe which was used at an unknown observatory is described, see figure 2.

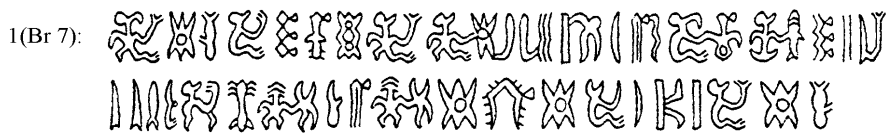

Figure 2.

---

[4] Smith, Carlyle S., 1962. An Outline of Easter Island Archreology. *Asian Perspectives*, 6, 1-2, p. 242.
[5] The studies are based on my own classification and translation scheme in deciphering the *rongorongo* signs: Rjabchikov, Sergei V., 1987. Progress Report on the Decipherment of the Easter Island Writing System. *Journal of the Polynesian Society*, 96(3), pp. 362-363, figure 1; Rjabchikov, Sergei V., 1993. Rapanuyskie texty (k probleme rasshifrovki). *Etnograficheskoe obozrenie*, 4, 126-127, figure 1; Rjabchikov, Sergei V., 1994. *Tayny ostrova Paskhi*, vol. 3. Krasnodar: Ecoinvest, p. 3, figure 1. I always take into account the vocabularies and rules of alternating sounds of the Polynesian languages, cf.: Tregear, Edward, 1891. *The Maori-Polynesian Comparative Dictionary*. Wellington: Lyon and Blair. pp. XIV-XXIV.
[6] Lee, Georgia, 1992. Op. cit., p. 145, figure 5.15.
[7] Rjabchikov, Sergei, 2013. Op. cit., p. 5, figure 6.
[8] The nomenclature and tracings of the Rapanui classical inscriptions are taken from this edition: Barthel, Thomas S., 1958. *Grundlagen zur Entzifferung der Osterinselschrift*. Hamburg: Cram, de Gruyter. The records on the New York figurine are designated as text X.
[9] Rjabchikov, Sergei, 2013. Op. cit., p. 7. See also: Rjabchikov, Sergei V., 2010. Rapanuyskaya observatoriya v Mataveri. *Visnik Mizhnarodnogo doslidnogo tsentru "Lyudina: mova, kul'tura, piznannya"*, 27(4), pp. 66-75.

1 (Br 7): **6/44 7 9 6 17 132 7-7 6 6-7 4 4-33 4 44b 3 26 50 44-68 6/3 11 52 4 5-15 5 5-12 26 6-15 118 6-33 84 65 3 26 6-33 84 7 74 7 6 3 4-41 4-6 7 9** *Hata Tuu Niva atea. Kore Tuutuu [Tuu Pipiri or Piri] a Hatu atua atua/ua. Ti Tua Hina maa, hi. Tahonga a Hina Mango hiti ti. Atua roa titika ma Hora, paka. Hau IVI RANGI Hina maa, hau IVI Tuu tini, Tuu a hina Tireo tuha, Tuu Niva.* 'The bright star from the Darkness (Grave, Winter) rose. Canopus disappeared in the rays of the rising sun. The stick – the settings of the bright (full) moon before the sunrises. The priest (expert) *Hina Mango* lifted the stick. The great lord marked lines for the month *Hora*, the dry season. (They were) the ropes (= lines) of the bright moon, the ropes (= lines) of the star (Aldebaran) on the day of the solstice, of (this) star on the night of *Tireo*, of the star from the Darkness (i.e., Canopus).'

In the script the star *Pipiri* (Canopus) can be designated with the glyph combination **7-7** (as a bright star, cf. Rapanui *piri* 'to join').[10] The full name of the star was **7-7 5** *Tuutuu [Tuu Pipiri or Piri] Atua* indeed. Maori *Atutahi* 'Canopus' came from the words *Atu(a) Tahi* 'The 1st God.' Old Rapanui *hata* means 'to rise; to elevate,' cf. Maori *whata* 'to elevate.' This term is preserved in the Rapanui place name *Hanga Tuu hata* 'The Bay 'The star is rising'.' Looking from the environs of the Mataveri area, Aldebaran rose at that place in the morning (for instance, in June, A.D. 1600).

The following fragments are interesting in the folklore text "Apai":
(1) *Ka Pipiri te hetuu tau avanga.* '(It is) the star Canopus of the time of the grave (= the winter).'
(2) *Ko hao ko Piri e Atua.* '(The star) *Piri e Atua* (*Pipiri* = Canopus) rose.'
(3) *Ko hao ko Piri e uta.* '(The star) *Piri* (*Pipiri* = Canopus) rose.'
(4) *A tara. Ka hiri a Uka hopua.* '(It was) the beginning of the morning. The washing Canopus (associated with rains) rose.' Cf. Maori *tara* 'rays of the sun, shafts of light, appearing before sunrise.'[11]

**The Great Astronomer *Hina Mango***

Glyphs **6/3 11** *a Hina Mango* in the read inscription (Br 7) are a real quasi-bilingual source for the decipherment, see figure 3.

1(Br 7): 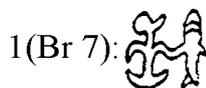

Figure 3.

Glyphs **6** *(h)a* denote particles of the personal names (*a*) in the famous genealogy on the Small Santiago tablet (Gv 5-6).[12] Glyph **3** *hina* represents the lunar crescent (cf. Rapanui *Hina* 'the moon goddess'). Glyph **11** *mango* represents the shark (cf. Rapanui *mango* 'shark').

An exact expression, *a Hina Mango*, is presented in a Rapanui song.[13] It is obvious that this text was a didactic source in the royal *rongorongo* school.[14] In the beginning of the chant it is said that somebody sent a *rongorongo* message to *Hina Mango* about King *Kai Makoi* [the First]: *teke te Makoi* 'King (*teketeke*) (*Kai*) *Makoi*' and about the harvest of the *niu* nuts: *niu hau pu*. It is of first importance that the answer of *Hina Mango* (cf. the words: *he unga e te Manga* [i.e., *Mango*]) contains a lot of astronomical details:
(1) *A Uka a tea.* 'Canopus shone.'
(2) *Tataki po ihuihu.* '(It was) the counting of the nights of the narrow lunar crescent.' (Cf. Rapanui *taku* 'to predict,' *tataku* 'to add; to count,' *ihu* 'nose.')

---

[10] Barthel, Thomas S., 1958. Op. cit., p. 218.
[11] Rjabchikov, Sergei V., 2011. Canopus and the Pleiades in Records on the Tahua Tablet. *Polynesian Research*, 2(1), pp. 11-12.
[12] Rjabchikov, Sergei V., 1988. Note on Butinov and Knorozov's Investigation. *Rapa Nui Journal*, 2(2), p. 6.
[13] Blixen, Olaf, 1979. Figuras de hilo tradicionales de la Isla de Pascua y sus correspondentes salmodias. *Moana: Estudios de Antropologia Oceania*, 2(1), pp. 74-78.
[14] See this paper about the local schools: Rjabchikov, Sergei V., 2012. The rongorongo Schools on Easter Island. *Anthropos*, 107(2), pp. 564-570.

(3) *Tataki po aveave*. '(It was) the counting of the nights of the full moon.' (Cf. Rapanui *avae* 'the full moon; lunar month,' Old Rapanui *ava* 'ditto.')
(4) *He ruru peaha, he kena peaha, ha hei peaha*. '(It was the determination of the exact time:) Perhaps the boobies (*ruru*) (appeared), perhaps the boobies (*kena*) (appeared), perhaps those that had been driven (*ha hei*, the manu-tara birds, figuratively) (appeared).'
(5) *Maharu; te Uka*. '(It was) the 11th night *Maharu*; Canopus (was seen).'
(6) *Ngutungutu po*. '(They were) the nights during the bird (months *Hora*, August and Seprember chiefly).' (Cf. Rapanui *ngutu* 'beak.')
(7) *Ko reva, ko reva*. '(It was) the height, (it was) the height (= some of the sun's positions).' (Cf. Maori *rewa* 'height.')
(8) *Ka koekoe mai te more o te Nuahine. More ka tatau, ka tatau ra. More ka rereva, ka rereva ra*. (It was the assertion about a solar eclipse: cf. Rapanui *koekoe* = *korekore* (absence), *more* (to break), *Nuahine* = *Hina*; cf. also *tau* 'time' and *raa* 'the sun.') Here the total solar eclipse of **September 16, A.D. 1773** (the 1st day, *Hiro* of the 4th month, *Hora-nui*) is described.

    The final part of the song was devoted to the recording of these letters on school lessons: *He tukia e te hahatu. Ka oho mahaki a potu e. Pu ti, pu ta hava, hava re*. '(It was) the inscription (*tuki*, *tuku*; *hahatu* = *hati*). Pupils (companions literally) who had been removed (*patu*) entered (the schoolhouse). All (the children) wrote (*ti*, *ta*) (this text) during 2 months (*ava*, *avae*) quickly (*rere*).'

    Thus, the great Rapanui astronomer *Hina Mango* named after the moon and belonged to the Miru tribe lived at the end of the 18th century A.D. I suppose that he could predict lunar and solar eclipses. Perhaps he created the stone calendar at Mataveri or an analogous device.

**Astronomical Simulations as the Main Clue: About other Petroglyphs**

On the wall of one house of the Orongo village a complicated plot is depicted.[15] Here the lizard glyph **69** *moko* and glyph **1** *Tiki* resembling a round (the solar symbolism) are united. The style of the signs is not archaic. The calculations have been carried out for the 4th lunar month (beginning in September in the majority of instances) from A.D. 1700 till A.D 1800. It may be the description of the almost total solar eclipse of **August 24, A.D. 1710** or of the total solar eclipse of **September 16, A.D. 1773** (in both cases it was the 1st day, *Hiro* of the 4th month, *Hora-nui*).

**Decoding the Script: Astronomical and Calendar Data**

Let us examine the following fragments on the tablets Tahua (A), Aruku-Kurenga (B), the Santiago staff (I), the Great St. Petersburg (P) and the Great Santiago (H) tablets, see figure 4.

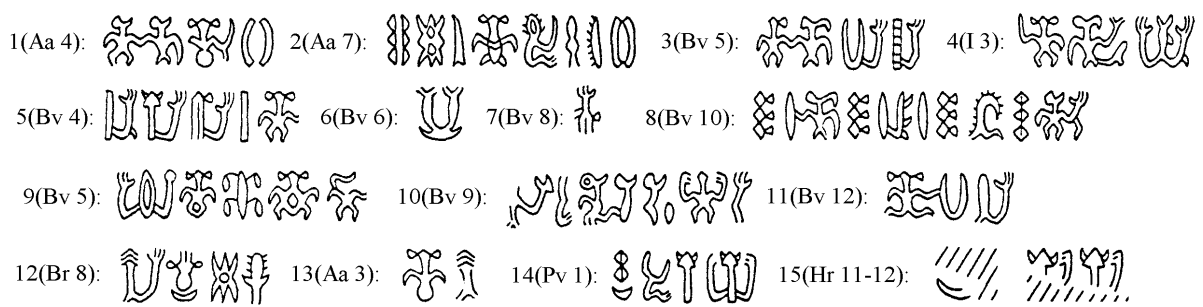

Figure 4.

---

[15] Koll, Robert R., 1991. Petroglyphs inside Orongo's Houses, Easter Island. *Rapa Nui Journal*, 5(4), p. 62, figure. See the interpretation in Rjabchikov, Sergei V., 1997 *Tayny ostrova Paskhi*. Vol. 6. Krasnodar: The Chamber of Commerce and Industry of Krasnodar Region, p. 23; Rjabchikov, Sergei V., 1997. Rongorongo versus Kai-kai: A Second Look at Themes Linking Easter Island's Mysterious Script with Its String Figure Repertoire. *Bulletin of the International String Figure Association* 4, p. 37.

1 (Aa 4): **6-44 68 5 57** *Hata Ono atua, tara*. 'The deity 'The Pleiades' rose, (then) the rays of the sun before the sunrise (appeared).'
2 (Aa 7): **17-17 7-7 5 68 2 14 73-46 57** *Teatea Tuutuu* [*Tuu Piri* or *Pipiri*] *Atua, Ono, Hina Haua, Hena* (*Henga*) *tara*. 'Canopus, the Pleiades, the moon *Haua* (*Atua*) shone, (then) the sunbeams (appeared).'

For both records choose, as a case in point, June 23, A.D. 1600. The azimuth of the setting moon (4:33) of the phase *Haua* (*Atua*) was 243°24'15". Canopus (SE), the Pleiades (NE), and the moon were seen (4:25) before the dawn (05:43).

In a Rapanui folklore text known as the Creation Chant the following sentence is presented: *Matua-Anua ki ai ki roto ki a Pipiri-hai-tau, ka pu te miro*.[16] In my opinion, this text can be translated as follows: '*Matua-Anua* (= the first Rapanui King *Hotu Matua*) by copulating with the star Canopus (beginning this) time produced the canoe (ship).'

*Anua-motua* was the name of a Mangarevan voyager who set out in a double canoe to the island Mata-ki-te-rangi (Easter Island).[17] Canopus could be the star of the navigation.
3 (Bv 5): **6-44 61 15 4-15 …** *Hata Hina roa atua roa…* 'The great goddess 'The Great Moon' rose…'
4 (I 3): **6-44 (102) 53/(123)** *Hata Maru* (*Maro*). '(The crescent) of (the 1st month) *Maru* (*Maro*) rose.'

*Maro* (*Maru*) (June chiefly) was the 1st month of the year in the past.[18]
5 (Bv 4): **4-15 21-15 26-15 4-6** *Atua roa Koro, Maro tuha*. 'The great deity 'The month *Koro* (December chiefly),' 'The month *Maro* (*Maru*; June chiefly),' the interval.'

Old Rapanui *tuha* 'interval' correlates with Rapanui *tuha* 'part; to divide.'
6 (Bv 6): **3 53** *HINA Maru* (*Maro*). 'The month *Maru* (*Maro*).'
7 (Bv 8): **26-15** *Maro*. '(The month) *Maro* (*Maru*).'

Here the name of the month is written down with two quasi-syllables, *ma* and *ro*.
8 (Bv 10):**17 30-44 17 30 51 30 17 14 17 6-15** *te Anakena, te Anakena, te Haua, te Hora*. '(The month) *Anakena* (July chiefly), (the night/day) *Haua*, (the month) *Hora* (*Hora-iti*, August chiefly).'

In this text the name of the 2nd month *Anakena* was taken down twice, as **30-44** *Ana-kena* and **30 51 30** *Ana-ke-ᵃna*.
9 (Bv 5): **15-28 4 68 25 6-99** *Rongo-atua hono hua hami*. 'The god *Rongo* (of the Orongo village) added the egg in the morning.'

The site Orongo was named after the god *Rongo*.[19] Old Rapanui *hami* 'dawn; to dawn' corresponds to Rapanui *hamu* 'to dawn.'[20] *Te Amira* was a guardian spirit at the quarry Puna Pau.[21] His personal name reads *Ami raa* 'The dawn.'
10 (Bv 9): **31 43 19 27-27 30 6-15** *MAKE Maa-ki raurau ana Hora*. '(The god) *Makemake* created the shine (or abundance) of the month *Hora*.'

In this record the variant of the *Makemake*'s name, *Maki* (*Maa ki*), has the same structure as the name *Tiki* (< PPN *Tii ki*).
11 (Bv 12): **69 61 48-15** *Moko: hina uri*. 'The Lizard (the night/day *Hiro*): the black moon (= the absence of the moon in the sky).'
12 (Br 8): **4/33 3 56-32 7 25 …** *Atua roa hina, Pou, Tuu Hu…* 'The great goddess 'The Moon,' Sirius, Aldebaran…'
13 (Aa 3): **68 33/59** *Ono; Uka*. 'The Pleiades; Canopus.'
14 (Pv 1): **3 17 2 21-15-21-15** *Hina tea, HINA Korokoro*. 'The bright moon; the month *Koro* (December chiefly).'

---

[16] Métraux, Alfred, 1940. Ethnology of Easter Island. *Bishop Museum Bulletin 160*. Honolulu: Bernice P. Bishop Museum, p. 320.
[17] Buck, Peter H. (Te Rangi Hiroa), 1938. *Vikings of the Sunrise*. Philadelphia – New York, J.B. Lippincott Company, p. 207.
[18] Rjabchikov, Sergei V., 1993. Op. cit., p. 134.
[19] Best, Elsdon, 1924. *The Maori*. Vol. 1. *Memoirs of the Polynesian Society*. Vol. 5. Wellington: The Polynesian Society, p. 139.
[20] Rjabchikov, Sergei, 2013. Op. cit., p. 4.
[21] Barthel, Thomas S., 1978. *The Eighth Land. The Polynesian Discovery and Settlement of Easter Island.* Honolulu: University of Hawaii Press, p. 264.

15 (Hr 11-12): **3 … 21-15-21-15** *Hina … Korokoro.* '[The bright] moon; [the month] *Koro* (December chiefly).'

Fragments 14 and 15 contain parallel texts. *Koro* was the month of the summer solstice.

Liller thinks that the following plot of the local rock art is relevant to stellar motifs,[22] see figure 5.

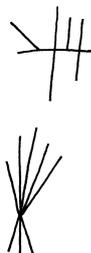

Figure 5.

This text written down in the cursive style reads thus: **7 26-15** *Tuu Maro.* 'The star of (the month) *Maro* (*Maru*) = Aldebaran.'

## The Terminology of Ancient Astronomers

On the strength of Mulloy's field notes Liller has published the list of some Rapanui ceremonial platforms that have their seaward facades aligned with the critical solar directions.[23]

The names of these platforms may be an additional source of the ancient astronomical terminology. They are as follows: *Rua Motu*, *Motu Rau*, *Vai Tara Kai Ua*, *Te A Kava*, *Parai A Ure*, *Motu Hitara*. The term *motu* in some names denotes the stone construction (cf. Rapanui *motu* 'to cut').

(1) The term *rua* could be the designation of the setting sun, cf. Maori *rua* 'setting-place of the sun.'

(2) The term *rau* (*rahu*, *rou*, *rohu*) denotes the increasing of the height of the sun above the horizon.

(3) The term *tara* is associated with the sunbeams before the sunrise (see above). The term *kai* (to eat) is relevant to eclipses.[24] Cf. Rapanui *vai* 'water' and *ua* 'rain'.

(4) According to natives' reports, the ghosts *Kava Aro* and *Kava Tua* lived in the regions of Maunga Parehe or *Raai*.[25] These gods were incarnations of the shining sun (cf. Rapanui *aro* 'face') and the setting sun (cf. Rapanui *tua* 'back').[26] The presence of two grammatical articles, *te* and *a*, one by one, has not been confirmed in folklore texts at all, so I prefer to read them as the word *tea* (to shine).

(5) The term *parai* (< *para*) is comparable with Maori *para* 'to shine clearly.' The term *ure* is associated with the fertility and abundance.

(6) The name *Hitara* consists of the archaic words *hi* (solar rays; to shine) and *tara* (solar rays before the sunrise), cf. Maori *hi* 'dawn' and *hihi* 'ray of the sun.'[27] Two names of a month, *Tarahau* and *Tarahao* (March chiefly; the month of the autumnal equinox), contain the same term *tara*.

---

[22] Liller, William, 1991. Hetu'u Rapanui: The Archaeoastronomy of Easter Island. In: Phyllis M. Lugger (ed.) *Asteroids to Quasars: A Symposium Honouring William Liller*. Cambridge: Cambridge University Press, p. 271, figure 2.

[23] Ibid., p. 277, table 2.

[24] Métraux, Alfred, 1940. Op. cit., p. 52.

[25] Englert, Sebastian, 1974. *La Tierra de Hotu Matu'a. Historia y etnologia de la Isla de Pascua*. Santiago de Chile: Ediciones de la Universidad de Chile, p. 137; Heyerdahl, Thor, 1976. *The Art of Easter Island*. London: George Allen & Unwin, p. 117.

[26] Rjabchikov, Sergei V., 2001. *Rongorongo* Glyphs Clarify Easter Island Rock Drawings. *Journal de la Société des Océanistes*, 113(2), p. 216.

[27] See my notes about the Old Rapanui forms *para* and *hi*: Rjabchikov, Sergei V., 1997. A Key to Mysterious Easter Island Place-Names. *Beiträge zur Namenforschung. Neue Folge*, 32(2), pp. 207-208.

## The Observatory at Tongariki. The Great Astronomer *Rahu* (*Rahi*)

The ceremonial platform Ahu Tongariki is situated near the southern coast of the island. In conformity with Mulloy and Liller, the facade perpendicular of this platform has the orientation to the summer solstice.[30]

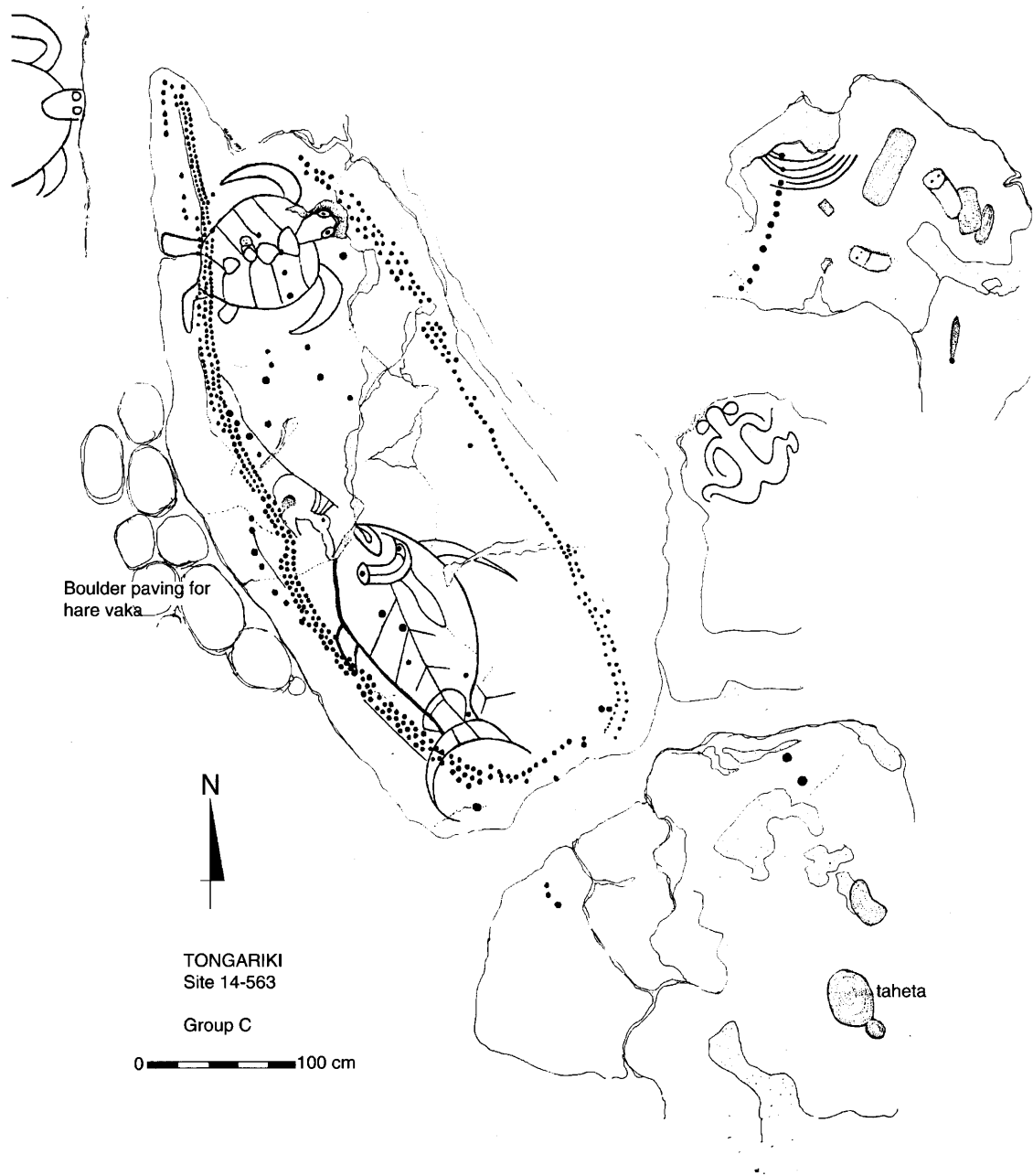

Figure 6.

This platform was dedicated to the sun deity. Actually, its stone construction is decorated with the grand glyph **1** *Tiki*.[31] According to a legend, during the drought a priest asked this god to hide his face.

---

[30] Liller, William, 1991. Op. cit., p. 284, table 3.
[31] Rjabchikov, Sergei V., 2010. Rapanuyskie zhretsy-astronomy v Tongariki. *Visnik Mizhnarodnogo doslidnogo tsentru "Lyudina: mova, kul'tura, piznannya"*, 24(1), p. 69.

Then he asked the deity *Hiva Kara Rere* to invoke rain.[32] Hence, the disappearance of *Tiki*'s face corresponds to the darkness of the sun and the heavens (rains; solar eclipses).

Let us study a fragment of a chant including some words about the platform Tongariki: *Te kere, mea, te kere, mea. Ko piti ko pata. Ko houhou parera aiai. Mata puku, Mata revareva. Ko Tonga, ko Tongariki. I ngaro ro ai.*[33] In my opinion, this text can be translated as follows: 'The black colour, the red colour (were seen), the black colour, the red colour (were seen) [= solar eclipses]. (The sun) was removed (*patu*) from the ecliptic (*pito*). The lines (*hau*, ropes) were from the shore [the sea bottom literally] at this place. The Face/Eyes (the sun) was at the height, the Face/Eyes (the sun) was at the height. (It was the place) Tonga (or) Tongariki. The eclipse (occurred).'

Here Old Rapanui *piti* 'ecliptic' (< \**pito*) corresponds to Maori *pito* 'ditto.'[34]

Let us investigate some elements of the scene presented in figure 6.[35] It is common knowledge that in the Rapanui mythology the tuna fish was the symbol of the god *Tangaroa*.[36] In this picture the tuna fish is *Tangaroa*. According to a legend, this deity (king) as a seal once went to Easter Island, arrived to Tongariki, and the people tried to roast him later.[37] His brother, *Teko*-of-the-Long-Feet (the god *Rongo* according to Fedorova),[38] arrived to the island to seek him. In this picture the god *Rongo* is shown as the turtle moving from the legendary western land (Hiva). Two turtles transmit *Rongo*'s motion indeed.

In Western Polynesia *Tagaloa* (*Tangaroa*) is the sun deity.[39] The story of his captivity is the figurative description of solar eclipses. The name *Tongariki* reads *Tonga Ariki* 'The Setting of the King [*Tangaroa*; the sun]' in fact, cf. Maori *tonga* (< *to-nga*) *o te ra* 'sunset.'

Consider the following record on the Aruku-Kurenga tablet (B), see figure 7.

1(Bv 7-8): 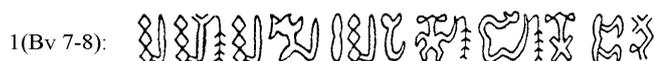

Figure 7.

1 (Bv 7-8): **17-4 17 15-24 17-4 3 84 4 65 17-4 27 6 24 62 24 49 50 29 70** … *Te atua tea ro ai, te atua hina, ivi atua rangi, te atua Rau a ai To(nga) ari(k)i (ariki) mau. Hi rua, pua…* 'The lord observed, (it was) the lord (= the observer) of the moon, the priest (= the observer) of the sky, (it was) the lord *Rau* (*Rahu*) of the site Tongariki (Tonga Ariki). The rays of the sun (were visible) during the settings (and) the risings…'

Old Rapanui *tea* (\**kitea*, cf. the metathesis, *tikea*) 'to see' is comparable with Maori *kitea* 'ditto.'

In conformity with Routledge, a scribe *Arahi* lived at Tongariki in the past.[40] This name includes the particle *a* and as such the proper name *Rahi* < \**Rahu*. According to the decoded text, that man was the great astronomer, too.

One can attempt to search for astronomical and calendar records at the Tongariki site. First of all, consider signs on another Tongariki's panel.[41] Here the tuna fish and the turtle are depicted as well, in the other words, they are the deities *Tangaroa* and *Rongo* (*Teko*). Two faces (glyphs **60** *mata*) are the sym-bols of the sun. The 30 dots (cupules) in 10 rows of 3 each are depicted on the right. I suppose that it is the record of the standard duration of the lunar month (30 nights/days).

---

[32] Felbermayer, Fritz, 1963. Hiva Kara Rere, der Gott des Regens. *Tribus*, 12, pp. 215-218.

[33] Barthel, Thomas S., 1962. Rezitationen von der Osterinsel. *Anthropos*, 55(5/6), p. 847.

[34] Best, Elsdon, 1922. The Astronomical Knowledge of the Maori. *Dominion Museum Monograph No 3*. Wellington: W.A.G. Skinner, Government Printer, p. 12.

[35] Lee, Georgia, 1992. Op. cit., p. 127, figure 4.134.

[36] Fedorova, Irina K., 1978. *Mify, predaniya i legendy ostrova Paskhi*. Moscow: Nauka, p. 24.

[37] Métraux, Alfred, 1940. Op. cit., pp. 310-311.

[38] Fedorova, Irina K., 1978. Op. cit., p. 23.

[39] Polinskaya, Maria S., 1986. *Mify, predaniya i skazki Zapadnoy Polinezii (ostrova Samoa, Tonga, Niue i Rotuma)*. Moscow: Nauka, p. 131.

[40] Routledge, Katherine, 1914-1915. Katherine Routledge Papers. Royal Geographical Society, London, Archives. Copies Held at Auckland Public Library, Auckland, New Zealand; Pacific Manuscript Bureau, Australian National University, Canberra, Australia; Instituto de Estudios, Universidad de Chile, Santiago de Chile; Rock Art Archive, The Institute of Archaeology at UCLA, Los Angeles, USA.

[41] Lee, Georgia, 1992. Op. cit., p. 83, figure 4.65.

Now one can return to figure 6. The 5 crescents, 12 dots (cupules) and glyph **69** *Moko* (the night/day *Hiro*) are represented on the right. They denote 5 lunar months plus 13 nights/days, i.e. 163 nights/days.

This result is crucial for the purposes of this research. On the Mal'ta baton, a calendar device of the Old Stone Age of Siberia, which served for prediction of lunar and solar eclipses the following groups of special signs are discovered: 53, 33, 30, 23, 16, 7, i.e. 162 marks. If someone counted the signs since the night of the full moon (the lunar eclipse), he should have computed the day before the new moon.[42] Larichev insists that the Siberian astronomers of the Old Stone Age knew that at the same place a new solar eclipse could recur in 669 synodic months (the Great Solar Saros).[43]

The priests-astronomers of Tongariki could predict the total solar eclipse of **September 16, A.D. 1773** (the 1st day, *Hiro* of the 4th month, *Hora-nui*). This date minus 669 synodic months is equal to **August 15, A.D. 1719** (the 1st day, *Hiro* of the 3rd month, *Hora-iti*). A partial solar eclipse occurred on that day. Alternatively, a partial lunar eclipse occurred on the night of **April 7, A.D. 1773**. The total solar eclipse of **September 16, A.D. 1773** happened in 162 nights/days after that lunar eclipse, so that both eclipses encompassed 163 nights/days. Thus, it is highly plausible that on the Tongariki panel these events are shown. What is more, the same interval of 163 nights/days had the almost total solar eclipse of **June 9, A.D. 1592** and the total lunar eclipse of **December 29, A.D. 1591**.

A report from the Tongariki observatory about a solar eclipse is presented on the Small Santiago tablet (G), see figure 8.

1(Gv 8): 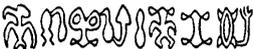

Figure 8.

1 (Gv 8): **44 73 25 60 25 22 65 6 100-100 28 15 …** *Tae hua. Mata hua hao, rangi a nono; ngaro…* (It was) not the same (situation). The same Face (= the sun; the sun god) rose, (but) the sky was unusual; (so that) an eclipse (occurred)…'

**A Female Statue as the Moon's Image**

In compliance with Barthel, a female statue (*Moai* 287) on the slope of the Rano Raraku volcano looks in a WSW. direction (approximately 250°).[44]

I suggest that the monument represents the full or almost full moon in the month *Maru* (*Maro*; June chiefly). The azimuth was 243°24'15" for the moon in the 13th phase *Atua* (*Haua*) on June 23, A.D. 1600. The azimuths of the setting moon increased on the next nights.

According to the local mythology, the deity *Rua Haua* lived in the Rano Raraku crater.[45] This personal name means 'The setting of the moon *Haua*'. It can be the domestic name of that statue.

**Conclusions**

It is apparent that the archaeoastronomy of Easter Island is an important key to the mysteries of that great civilisation. The priests-astronomers watched, among others, the sun, the moon, β and α Centauri, Aldebaran, Canopus, and Sirius. The facts of the predictions of solar and lunar eclipses are beyond question.

---

[42] Larichev, Vitaly E., 1993. *Sotvorenie Vselennoy: Solntse, Luna i Nebesny drakon*. Novosibirsk: Nauka, pp. 144-147.
[43] Ibid., pp. 209, 223.
[44] Barthel, Thomas S., 1958. Female Stone Figures on Easter Island. *Journal of the Polynesian Society*, 67(3), p. 253.
[45] Métraux, Alfred, 1940. Op. cit., p. 383.